\documentclass[runningheads]{comsis2}
\usepackage{graphicx}
\usepackage{tabularx}
\usepackage{supertabular}
\usepackage{amsmath}
\usepackage{txfonts}

\usepackage{graphicx}
\usepackage{amssymb}
\usepackage{amsmath}
\usepackage{algpseudocode}
\usepackage{algorithm}
\usepackage{multicol}
\usepackage{lineno}
\usepackage{booktabs}
\usepackage{lscape}
\def\journalissue{Computer Science and Information Systems ?(?):??--??}
\def\paperidnum{DOI: N/A}
\title{A GRASP approach for solving the 2-connected m-dominating set problem}

\titlerunning{A GRASP approach for solving the 2-connected m-dominating set problem}

\author{Raka Jovanovic \inst{1}         \and  Islam Safak Bayram \inst{1} \and 
         Stefan Vo{\ss}\inst{2}\inst{31}} 

\institute{Qatar Environment and Energy Research Institute (QEERI), Hamad bin Khalifa University,\\
  PO Box 5825, Doha, Qatar,\\
  \email{rjovanovic@qf.org.qa}
  \and
  Institute of Information Systems, University of Hamburg,\\
  Von-Melle-Park 5, 20146 Hamburg, Germany\\
  \email{stefan.voss@uni-hamburg.de}
  \and
  Escuela de Ingenieria Industrial, Pontificia Universidad Cat\'olica de Valpara\'iso, Chile\\
  }

\begin{document}

\maketitle

\begin{abstract}
In this paper, we present a constructive heuristic algorithm for  the  $2$-connected $m$-dominating set problem. It is based on a greedy heuristic in which a  2-connected subgraph is iteratively extended with suitable open ears. The growth procedure is an adaptation of the breadth-first-search which efficiently manages to find open ears. Further, a heuristic function is defined for selecting the best ear out of a list of candidates.  The performance of the basic approach is improved by adding a correction procedure which removes unnecessary nodes from a generated solution. Finally, randomization is included and the method is extended towards the  GRASP metaheuristic. In our computational experiments, we compare the performance of the proposed algorithm to recently published results and show that the method is highly competitive and especially suitable for dense graphs. 

\vspace{6pt}\textbf{Keywords:} dominating sets, 2-connected graphs, GRASP, fault tolerant.
\end{abstract}

\section{Introduction}

\begin{figure}[thb]
\centering
\includegraphics[width=0.80\textwidth]{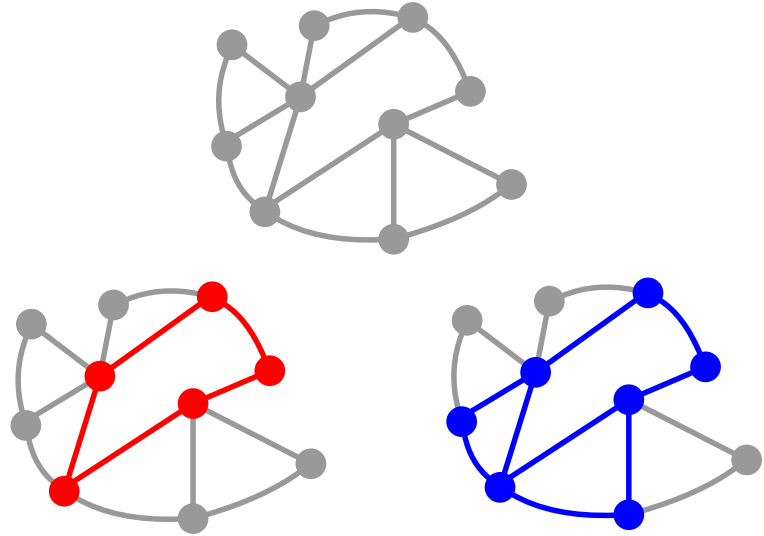}
\caption{Examples of a problem instance (top) and solutions for the 2-1-CDS (bottom left) and the 2-2-CDS (bottom right)}
\label{fig:Problem}
\end{figure}

A dominating set for a graph $G(V, E)$ is a subset of vertices $D \subseteq V$ that has the following property: every vertex in $G$ either belongs to $D$ or is adjacent to a vertex in $D$. Finding the dominating set with the smallest possible cardinality for a graph is one of the standard NP-hard problems \cite{du2012connected}. In this work, we focus on the $2$-connected $m$-dominating set  (2-$m$-CDS) problem which has two more constraints than the original problem. The first constraint is that there are two vertex disjoint paths between any two nodes $a, b \in D$  which are entirely within $D$ and the second one is that each node in $V \setminus D$ has at least $m$ neighbors in $D$. An illustration of a problem instance for the 2-$m$-CDS can be seen in Figure \ref{fig:Problem}.

A wide range of exact and approximate methods have been developed for finding solutions to the minimal dominating set problem (DSP). In the work of van Rooij et. al, to the best knowledge of the authors,  the fastest exact algorithm is presented which calculates the optimal solution in  $O(1.5048^n)$ time \cite{Rooij2009}. A variety  of metaheuristic approaches have  been developed for the DSP, some examples are the use of genetic algorithms \cite{Hedar2010} and ant colony optimization \cite{ACODSP}.    Several variations have been considered like the weighted \cite{jovanovic2010ant} and  connected version \cite{du2012connected} of the problem. The connected one  gives a more realistic model for many  real world systems. For it several different approaches have been used for finding either exact or approximate solutions \cite{du2012connected,RakaComSIS}.

In the recent years, there has been a growing interest in the DSP and its engineering applications. More specifically, DSPs are extensively used in wireless ad-hoc networks to develop routing algorithms and finding minimum size backbone in wireless networks \cite{du2012connected}. Therefore, the focus of this paper is on the application of the DSP for fault tolerant systems with particular emphasis in the field of wireless networks. The problem of the minimal $k$-connected $m$-dominating sets ($k$-$m$-CDS) has proven to be very suitable for modeling and optimization of such systems. There have been several different approaches for finding optimal and approximate solutions for the general problem and some of its restricted versions. Mixed integer programs (MIP) for the $k$-$m$-CDS have proven to be very successful in finding optimal solutions for medium-size graphs. In the work by Forte et al. \cite{do2013formulations} a MIP model has been defined for solving the $2$-$1$-CDS, and its performance is enhanced using a primal heuristic and Branch-and-Cut. Recently a very efficient MIP formulation based on vertex-cuts has been defined for the problem of interest, which is especially suitable for the $k$-$k$-CDS \cite{buchanan2015integer}. In the work of Ahn and Park, a MIP formulation for the $k$-$m$-CDS is proposed, and its performance is evaluated on a broad range of graphs \cite{Ahn2015}. 

On the other hand, research has also been conducted for finding approximate solutions for the $k$-$m$-CDS. In the paper by Daia and Wu, four localized $k$-CDS construction protocols are proposed \cite{dai2006constructing}. The relationship between the maximal independent set and the general $k$-$m$-CDS has been exploited for developing approximate centralized \cite{Thai200749,li2012construction} and distributed \cite{dai2006constructing,li2012construction} algorithms. In the work by Shi et al.  \cite{shi2016greedy} a greedy method is proposed for the 2-$m$-CDS for optimizing a fault-tolerant backbone of a wireless network. In their work, initially a 1-$m$-CDS is solved and later the desired connectivity is achieved through the merger of blocks. Authors have also explored the weighted version of the $k$-$m$-CDS for which they propose a constant approximation algorithm \cite{shi2015approximation}. Specialized constant approximation algorithms have also been developed for  3-$m$-CDS \cite{kim2010new,wang2013construction} for application in homogeneous wireless networks. A multiphase approximate method has been developed for the closely related 2-hop 2-connected dominating set \cite{zheng2012constructing}. Another related problem is the 2-$m$-domatic partition where the goal is to find some disjoint $r$-$m$-dominating sets in the network. For this problem, an approximate algorithm has been developed \cite{Jia2014}. 
 
In this paper, we propose a greedy heuristic method for solving the 2-$m$-CDS based on the existence of an open ear decomposition of a 2-connected graph \cite{EarDecomposition}. To be more precise, we iteratively grow a 2-connected subgraph $S$ by extending it with an open ear $P$. The growth procedure, which manages to find new open ears in an efficient way, is based on an adaptation of the breadth-first-search which has previously been successfully applied to the problem of maximal partitioning of graphs into 2-connected components with size constraints \cite{MBCPG}. The quality of found solutions is improved by adding a correction procedure which removes redundant nodes from the dominating set. Further, randomization is included and a Greedy randomized adaptive search procedure (GRASP) algorithm \cite{feo1995greedy} is developed. In our computational experiments, we show that the method is highly competitive with recently published research. 

The paper is organized as follows. In the following section we present the aforementioned growth procedure. In Section 3. we give an outline of the proposed greedy algorithm. In the next section details of the GRASP algorithm are presented. In the later section the results of the conducted computational experiments are shown. 
 
 \section{Growth Procedure}
 
\begin{figure}[thb]
\centering
\includegraphics[width=0.70\textwidth]{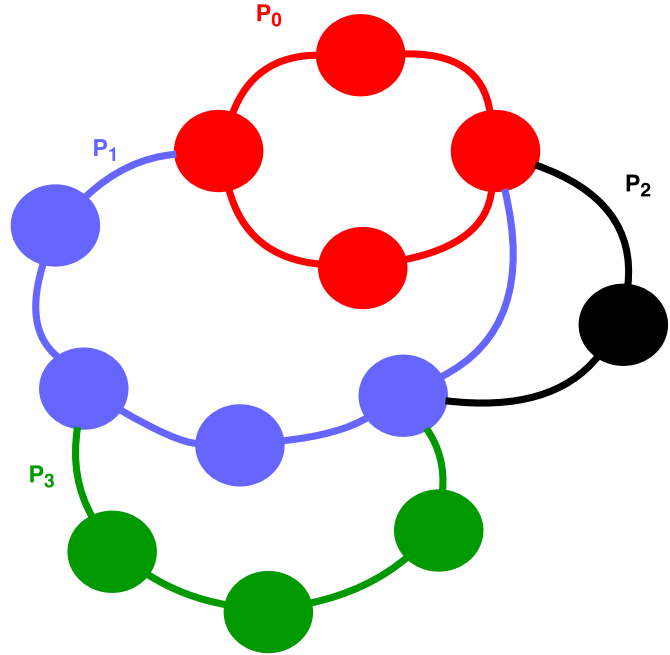}
\caption{Examples of an open ear decomposition for a bi-connected graph. Different colors are used for separate open ears.}
\label{fig:OpenEarDecomposition}
\end{figure}

In this section, we give a short outline of the procedure for growing a 2-connected subgraph, while the details can be found in \cite{MBCPG}. Since the proposed procedure is based on the fact that any 2-connected graph has an open ear decomposition we start with its definition. An ear of an undirected graph $G$ is a path $P$ where the two endpoints of the path may be the same, but no other edge or vertex appears more than once. In other words, any internal vertex of $P$ has degree two in $P$.  An open ear decomposition of graph G is a sequence of ears $P_0,\dots,P_n$ in which only $P_0$ is a cycle. This sequence must satisfy the constraint that the endpoints of ear $P_i$ belong to some $P_j, P_k$ where $j,k < i$ and $j,k$ are not necessarily distinct. No other vertices of $P_i$ can belong to any $P_j$ where $j<i$. An illustration of an open ear decomposition of a graph is given in Figure \ref{fig:OpenEarDecomposition}. 

The idea behind the method for growing a 2-connected subgraph $S$ is the following.  Starting from an initial cycle $S_0$  we iteratively generate a sequence of subgraphs  $S_0 \subset S_1 \subset S_2\subset \cdots$, where 
\begin{equation}
S_{i+1} = S_{i} \cup P_i 
\end{equation}
and $P_i$ is an open ear for $S_i$.  
It is noteworthy that due to the construction procedure each subgraph $S_i$ will have an open ear decomposition, hence, it will be 2-connected.  Such a sequence can easily be generated by adapting the BFS in a suitable way. 

The BFS is commonly used for finding cycles in the following way.  Let us assume that we start the BFS from some initial node $r$.
 As new nodes are visited and the BFS tree is expanded  the first time we encounter a back-edge $(u,v)$ and 
 an initial cycle $S$ is found. More precisely, the cycle consists of three segments: a BFS tree path from $r$ to $u$, the back edge $(u,v)$ and the BFS tree path from $v$ to $r$. As further nodes are visited and a new back-edge $(s,t)$
 is encountered a new open ear is found if some constraints are  satisfied. The notation $root(u,P)$ will be used for the node $v$ which is the first ancestor of $u$, in the BFS tree, such that $v \in P$. In case  $u \in P$,  $root(u,P) = u$.  For simplicity of notation we use $root(s)$ if $P$ is equal to the previously generated 2-connected subgraph $S$. Now, the constraints can be defined as: at least one of the nodes $s,t$ is not in $S$ and $root(s) \neq root(t)$. The new open ear $P$ will consist of the following segments: the BFS tree path from $root(s)$ to $s$, the back edge $(s,t)$ and the BFS tree path from $t$ to $root(t)$. It is obvious if we extend $S$ with $P$ we can repeat this procedure and further grow a 2-connected subgraph.

In the adaptation of BFS for growing a 2-connected graph, we explore the node $u$ having the minimal distance $d(u)$ to the already generated bi-connected subgraph $S$, instead of the distance to the root node $r$. The second change, compared to the original BFS, is that we need to track the values of $root(u)$, $d(u)$ for nodes in the BFS tree, which can change as a new ear $P$ is added to $S$. The notation $desc(u)$ will be used for the set of all descendants of $u$ in the BFS tree. We will also define its extension to node sets where $desc(P) = \bigcup_{i \in P} desc(u)$. The values of $root(u)$, $d(u)$ are calculated using an update procedure, performed after an ear $P$ is added to $S$, which is applied for all $u \in P \cup desc(P)$ in the following way.
\begin{eqnarray}
\label{CorrectRoot}
root(v) &=&   root(v,P).\\
\label{CorrectD}
d'(v) &=&   \left\{ \begin{array}{l}
 d'(v)- d'(root(v, P))\,\,\,\,\,\,\,\,\,\,\,\,\,\,\,\,\,\,\,\,\,\, v \notin P\\
 0\,\,\,\,\,\,\,\,\,\,\,\,\,\,\,\,\,\,\,\,\,\,\,\,\,\,\,\,\,\,\,\,\,\,\,\,\,\,\,\,\,\,\,\,\,\,\,\,\,\,\,\,\,\,\,\,\,\,\,\,\,\,\,\,\,\,\,\,\, v \in P\\
 \end{array} \right.
\end{eqnarray}
It is noteworthy that the proposed correction for functions $d(u)$ will produce approximations to the exact distance $d'(u)$. The function $d'$ satisfies that  $d'(u)\geq d(u)$ since it is possible to have an alternative path to $S$ using some back edges which is shorter. 

In the standard BFS, there is no change in the distance for visited nodes and no node is re-visited. In the proposed adaptation of BFS, such changes can occur and some revisits are necessary. The revisits are needed since some back edges that did not create open ears, may do so after the changes of the root values.  Both of these issues are addressed simultaneously using the following approach. First, nodes will be re-added to the queue as a new ear is added to $S$ and their re-evaluation is needed. Since it is possible for the same node to be added multiple times to the queue due to the addition of multiple ears, an additional value will be used to track if an evaluation is needed.  The algorithm for growing a bi-connected subgraph is better understood by observing Algorithm \ref{Alg:Grow}.
\begin{algorithm}
\begin{algorithmic}
\Procedure{BFSGrowBiConnected} {G, r}
\State{For all $u \in G$  initialize $Dist, Eval, Parent$ }
\State{Perform initialization for  $r \cup N(R)$ }
\State{Add all $N(r)$ to $Q$ }
\While{($Q$ is not empty) $\wedge$ ($NotExitCriteria$)}
 \State{$current = Q.dequeue()$} \Comment{Using Queue (FIFO) structure}
 \If{$current.Eval$}
 \ForAll{$u \in N(current)$}
 \If{$(u,current)$ is BackEdge)}
 	\If {$u,current$ produce an open ear $P$ }
 		\State{$S = S \cup P$} 
	 	\State{Perform necessary updates based on $P$} 
 	\EndIf
 \Else
 	\State{$u.[Root, Dist]= [current.Root, current.Dist+1]$} 
 	\State{Update parent, child relations for $u$, $current$ } 
	\State{$Q.enqueue(u)$ } 
 \EndIf
 \EndFor
  \State{$current.Eval = false$}
  \EndIf
\EndWhile
\EndProcedure
\end{algorithmic}
\caption{\label{Alg:Grow} Pseudo code for growing a bi-connected subgraph}
\end{algorithm}

\begin{figure}[!htb]
\centering
\includegraphics[width=1\textwidth]{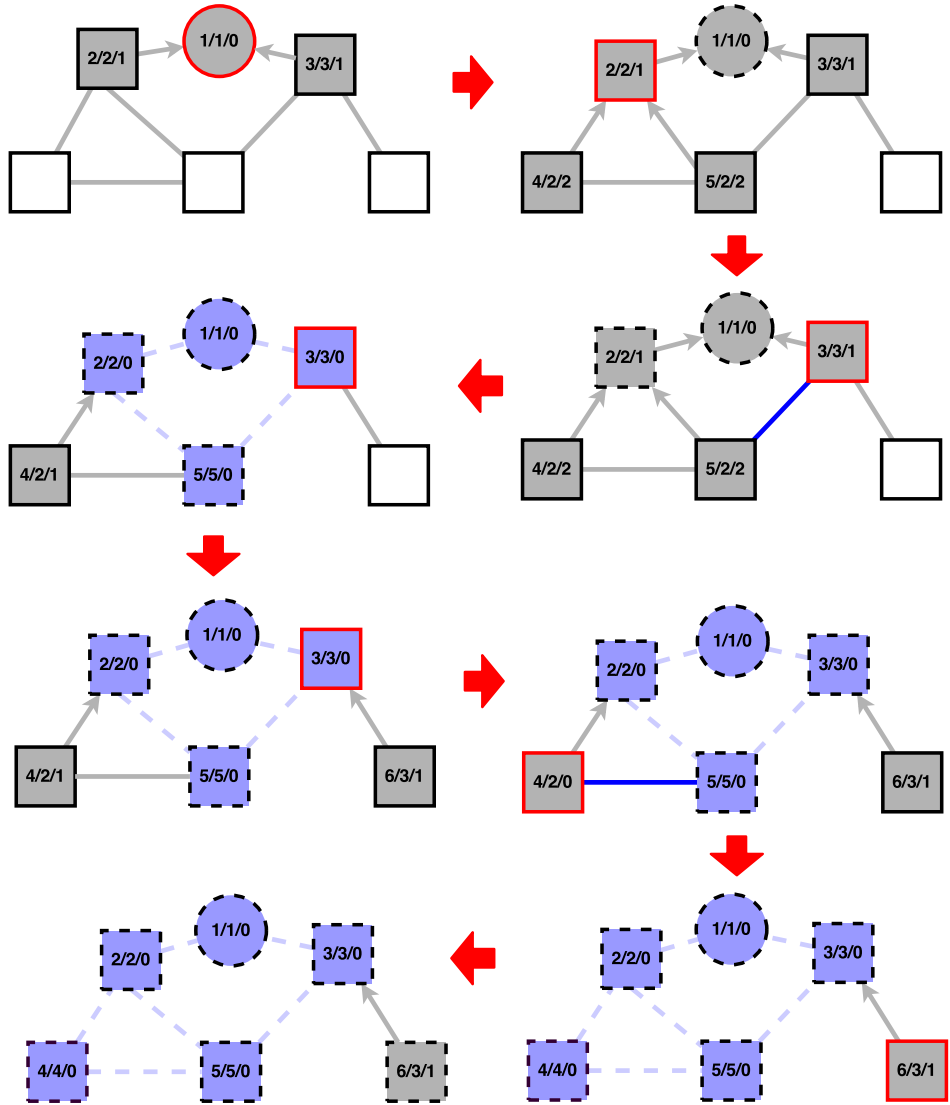}
\caption{Illustration of steps in the growth procedure. The root node is represented with a circle and the rest of the nodes are shown as squares. Each node $u$ stores 3 values, first is the id acquired by the BFS, the second is the value of $root(u)$ and the last is the values of $d(u)$. The node currently explored by the BFS is colored red. Gray arrows are used to indicate the structure of the BFS tree. Blue color edges are used for found back-edges at some step of the procedure. The violet color is used to show nodes that are a part of the found be bi-connected subgraph $S$ and the corresponding edges are presented using violet dashed lines. Dashed borders of a node $u$ indicate that the exploration of node $u$ has been completed.}
\label{fig:NormalStep}
\end{figure}

The proposed algorithm starts with a standard BFS initialization of the distance, parent and descendant relations for all the nodes with the additional property of the need for evaluation. Initially the evaluation property $Eval$ will be set to $true$ for all nodes. An auxiliary structure is used to store all the properties of individual nodes, which can be accessed and updated using the node id. Some special initialization is needed for the root $r$ and all its neighbors $N(r)$ for which details can be found in \cite{MBCPG}. Next, all nodes in $N(r)$ are added to the queue $Q$. The main loop is executed for each node $current$ in $Q$ until $Q = \emptyset$.  For each such node, we first check if an evaluation is needed and if so all its neighbors $N(current)$ are evaluated. For each $u \in N(current)$ we check if $(current, u)$ is a back-edge. In case it is not we add $u$ to the $Q$ as in the BFS, and we set $root(u) = root(current)$. In case $(current,u)$ is a back-edge we check if it induces a new open ear $P$ connected to $S$. If this is true the subgraph $S$ is extended with $P$ and necessary updates are performed. To be more precise, for all $u \in P$ the distance is set to $d.Dist = 0$. Each node $u$ now becomes a root of a new potential ear, so we set $u.root = u$.  For each node $u \in P$ we wish to update the branch of the BFS tree whose root is $u$. All such nodes  $v$ are added to $Q$ for re-evaluation and the values $v.root$, $v.Dist$ are corrected based on Eqs. \eqref{CorrectRoot},\eqref{CorrectD}.  The implementation details can be found in \cite{MBCPG}.   After all the elements of $N(current)$ are visited the evaluation of node $current$ is complete and we set $current.Eval = false$.  A graphical illustration of the basic steps of the growth procedure is given in Figure \ref{fig:NormalStep}.

 \section{Outline of the greedy algorithm}
 \label{GreSec}
 
The growth procedure presented in the previous section gives us an efficient way to grow a 2-connected subgraph by extending it with new open ears. Note that in practice it consists of two parts. The first one is finding a new ear and the second is adding an open ear to the already generated subgraph $S$ and performing updates to the BFS tree. The idea of the greedy procedure is to extend $S$ not with the first found open ear but with the best one, based on some heuristic function, from a suitable list of candidates. Let us first define the heuristic function for selecting an ear $P$ from a list of candidates. Our goal is to make a minimal dominating set, so there is a preference for expanding $S$ with a small ear whose nodes are adjacent to the highest number of not already dominated nodes. In the case of 2-$m$-DSP we will say that node $u$ is dominated by $S$ if it is adjacent to at least $m$  nodes in $S$. Let us define functions $\hat{dom}(u,A)$/$Cov(u,A)$ for a node $u$ and a set of nodes $A$ as the number/set of nodes $v \in A$ for which $(u,v) \in E$.  To that end, let us define 
\begin{equation} 
	\label{HeurNode}
   dom(u,A) =   \left\{ \begin{array}{l}
             min(\hat{dom}(u,A),m - dom(u,S) )  \qquad u\notin A  \\
             m - dom(u,S)  \qquad\qquad\qquad\quad \,\, u \in A   \\
 \end{array} \right.
\end{equation} 
The function $dom(u,A)$, in case of $u \notin A$, is a correction of $\hat{dom}(u,A)$ that has a limit that corresponds to the number of additional nodes that node $u$ needs to be adjacent to, so it is dominated by $S$.  Since $dom(u)$ is used for defining the heuristic function for selecting open ears, we also consider nodes $v \in A$ which in the proposed application become a part of the 2-connected subgraph. For such nodes, $dom(u)$ will be equal to $m - dom(u,S)$ since this could be understood as the additional number of adjacent nodes that are needed for $u$ to become dominated.
Using Eq. \eqref{HeurNode} we can define a heuristic function for candidate ears as  
 \begin{equation} 
	\label{HeurEar}
	dom(P) = \frac{\sum_{i \in \tilde{N}(P)} d(i,In(P))}{|In(P)|}  \qquad, \tilde{N} = N[In(P] \setminus (S)
\end{equation} 
In Eq. \eqref{HeurEar} the notation $In(P)$ is used for the set of inner nodes of $P$ , and $|In(P)|$ for the corresponding number of nodes. $N[P]$ is used for the set of all nodes $u$ in $P$ or adjacent to some node $v \in P$. The heuristic function $dom(P)$ is equal to the sum of $dom(u)$ for all inner nodes of $In(P)$ and corresponding neighboring nodes that are not already in $S$. The heuristic has a preference for ears that are adjacent to a large number of nodes not dominated by $S$ and have a low number of internal nodes. 

In the growth procedure presented in Algorithm \ref{Alg:Grow}, as soon as a new open ear $P$ is found, the subgraph $S$ is extended accordingly. In practice this is not necessary, but instead we can define a candidate list $C= \{ P_1, P_2, \cdots, P_M\}$ to which we add $P$. When there is a sufficient number $M$ of candidates in $C$ we can select the best ear $P_b$ based on the heuristic function $dom(P)$, and only then expand $S$ and perform the necessary updates to the BFS tree. In the implementation of the candidate list, it is most efficient to calculate the value $dom(P)$ and store the set of the nodes $U(P)$ whose elements are in $N[In(P)]$ that are not dominated by $S$. After the best (based on the heuristic) ear $P_b$ is added to $S$, the candidate list $C $ needs to be updated. The first update is to correct the values of $dom(P_i)$ and $U(P_i)$ based on the new ear $P_b$. This can be simply done by exploiting the stored values of $U(P_b)$, $U(P_i)$. The second update is for candidate ears for which $In(P_i) \cap In(P_b) \neq \emptyset$. For simplicity let us assume that there is only one node $u \in P_b \cap P_i$, and that  ear $P_i$ corresponds to some path $( a_0, \cdots, a_k, u, b_m, \cdots b_0)$. In such a case we must remove $P_i$ from the candidate list (since it is not a valid open ear) and add two new ones  $A = ( a_0, \cdots, a_k, u)$ and $B = ( u, b_m, \cdots, b_0)$ if they have a length of at least 3 nodes. The case when there are multiple nodes in $P_b \cap P_i$ is treated in a similar way. In practical applications, the candidate list will only contain an ear $P$  if it satisfies $dom(P) > 0$. Note that although this improves the performance of the algorithm in some case it can result in the method not being able to produce a feasible solution. 

 \section{GRASP}
To improve the performance of the greedy algorithm presented in the previous section we extend it to the GRASP metaheuristic. To do so we need to define a randomization for the greedy algorithm and develop a local search. The randomization can be done trivially by initially selecting a random root node $r \in G$ and by defining a new heuristic function $dom_r(P) = \alpha dom(P)$, where $\alpha$ is a random variable from some interval $(1, \alpha_0)$. 
For a local search we will use a simple correction procedure in which we iteratively remove nodes that are not necessary. The term "necessary" is used for a node $u$ for which $S_u = S \setminus \{u\}$ is either not a dominating set of $G$  or the graph induced by $S_u$ is not 2-connected. All the elements of the set of nodes $Q$ that are necessary for maintaining dominance satisfy the following simple constraint.  Any node $v \in S$ for which there exists a node $u \in G \setminus S$ such that $(u,v) \in E$  and   $\hat{dom}(u) = m$ is necessary.  For all nodes in $u \in S \setminus Q$ we can simply check if the graph induced by $S_u$  is a 2-connected graph by using  Tarjan's linear time algorithm  \cite{hopcroft1973algorithm}. Note that in the practical implementation set $Q$ can be easily updated based on the node $u$ that is removed from $S$.

\begin{algorithm}
\begin{algorithmic}
\While{ Not Max Solutions Generated}
\State{Select random $r \in G$ for root}
\State{Initialize BFS for $r$}
\State{ $Candidates = \emptyset$ }
\While{$S$ not dominating set}
 \If{$|Candidates| \geq MinNumberOfCandidates$}
 \State{Select Best $P \in Candidates$ based on $d_r(P)$}
 \State{$S = S \cup P$}
 \State{Perform updates of $BFSTree$ and $Candidates$}
 \Else
 \State{$P$ = FindSingleEar($BFSTree$)}
 \State{Add $P$ to $Candidates$}
 \EndIf
\EndWhile
\State{Calculate $Q$  for $S$}
\Repeat
\ForAll{$u \in S \setminus Q$}
\If{$S_u$ is 2-connected}
\State{Update $Q$ based on $u$}
\State{ $S = S_u$}
\State{\textbf{break}}
\EndIf
\EndFor
\Until{No Node has been removed from $S$}
\State{Check if $S$ is the new best solution}
\EndWhile
\end{algorithmic}
\caption{\label{Alg:GRASP} Pseudo code for GRASP  for 2-$m$-DSP}
\end{algorithm}
Details of the proposed GRASP algorithm can be seen in the Algorithm \ref{Alg:GRASP}. In the main loop, at each iteration a random initial node $r$ is selected as the root of the adapted BFS for growing 2-connected subgraphs and all corresponding initialization is done. The next loop generates an initial solution through the following steps. It first checks if the number of candidate ears is sufficient. If this is true, the best one, based on the randomized heuristic function $dom_r$, $P_b$ is selected and used to expand the partial solution $S$. Furthermore, all necessary updates are performed on the BFS tree and the candidate list. Otherwise, in case more candidates are needed, one is acquired using function $FindSingleEar(BFSTree)$. This function further grows the BFS tree until a new ear is found and returns it. It is important to emphasize that it does expand $S$. After, a dominating set $S$ is found the first step is calculating the set of necessary nodes $Q$. This set is used in the following loop used for improving the solution. In it, nodes are removed one by one from $S$ until no node can be eliminated. After each node $u$ is removed from $S$, the set  $Q$ is updated.  Finally, we check if the corrected dominating set $S$ is the best found.
 \section{Results}
In this section, we present the results of the computational experiments used to evaluate the performance of the proposed method. The  GRASP algorithm is compared to the recent MIP method from \cite{buchanan2015integer}. The presented method is implemented in C\# using Microsoft Visual Studio 2015. The calculations have been done on a machine with Intel(R) Core(TM) i7-2630 QM CPU \@ 2.00 Ghz, 4GB of DDR3-1333 RAM, running on Microsoft Windows 7 Home Premium 64-bit. We note that calculations in the article \cite{buchanan2015integer}  were conducted on a Dell Precision WorkStation T7500 R machine with two Intel XeonR  E5620 2.40 GHz quad-core processors and 12 GB RAM. The comparison has been made on 41 random graphs having 30 - 200 nodes and edge densities between 5-70\%. The graph instances are the same as in \cite{buchanan2015integer} which have also been used in  \cite{Buchanan2014410}. For each of the graph instances we compare the computational time and quality of found solutions for the 2-1-CDS and 2-2-CDS for all of which the MIP method managed to find optimal solutions.
  \begin{table*}[htb]
\scriptsize
\center
\caption{\label{table:ComGRASPMIP21CDS}Comparison of the  $MIP$ method from \cite{buchanan2015integer}, to the proposed greedy algorithm combined with the correction procedure $GrC$ and  the GRASP for the 2-1-CDS. "-" is used for instances without a feasible solution and "*" in case $GrC$ did not generate it.  Underlined values indicate that GRASP did not find optimal solution. }
\begin{tabularx}{350pt}{X*{8}{c}}

\toprule
Dataset& \multicolumn{3}{c}{$Solution$}& \hspace{50pt} & GRASP Iter. \hspace{10pt}&\multicolumn{2}{c}{$Time[ms]$} & \\
		\cmidrule(r){2-4}\cmidrule(r){7-8}
       &   GrC& GRASP& MIP&& &GRASP &MIP \\
\midrule
v30\_d10 & 18 & 18 & 18 && 6&1&20\\
v30\_d20 & 9 & 8 & 8 & &24&26&40 \\
v30\_d30 & 7 & 5 & 5 & &5&3&20 \\
v30\_d50 & 3 & 3 & 3 & &3&2&10 \\
v30\_d70 & 3 & 3 & 3 & &1&0&50 \\
v50\_d10 & 15 & 14 & 14 && 54&63&50 \\
v50\_d20 & 8 & 7 & 7 & &2&2&40 \\
v50\_d30 & 6 & 5 & 5 & &8&15&90 \\
v50\_d50 & 3 & 3 & 3 & &17&28&80\\
v50\_d70 & 3 & 3 & 3 & &1&1&80 \\
v70\_d5 & 40 & 34 & 34 & &6108&1.3e4&10 \\
v70\_d10 & 20 & 14 & 14 & &1037&2.9e3&1.0e2\\
v70\_d20 & 8 & 8 & 8 & &13&35&1.3e2 \\
v70\_d30 & 5 & 5 & 5 & &24&78&2.1e2 \\
v70\_d50 & 4 & 3 & 3 & &36&1.0e2&80 \\
v70\_d70 & 3 & 3 & 3 & &1&2&1.6e2 \\
v100\_d5 & 35 & 28 & 28 & &4264&1.7e4&2.0e2\\
v100\_d10 & 16 & 14 & 14 & &1531&6.4e3&2.9e2 \\
v100\_d20 & 11 & 8 & 8 & &612&2.8e3&7.3e2 \\
v100\_d30 & 7 & 6 & 6 & &39&1.7e2&1.0e3 \\
v100\_d50 & 4 & 4 & 4 & &6&27&5.3e2 \\
v100\_d70 & 3 & 3 & 3 & &1&3&1.3e3 \\
v120\_d10 & 18 & 14 & 14 & &6240&3.0e4&6.0e2 \\
v120\_d20 & 10 & 8 & 8 & &13413&7.4e4&3.8e3 \\
v120\_d30 & 8 & 6 & 6 & &239&1.4e3&1.2e3 \\
v120\_d50 & 4 & 4 & 4 & &5&36&3.3e3 \\
v120\_d70 & 3 & 3 & 3 & &1&5&5.2e2 \\
v150\_d5 & 35 & \underline{30} & 28&& [590]&3.6e3&9.3e2 \\
v150\_d10 & 18 & 15 & 15 && 5127&3.2e4&3.9e3 \\
v150\_d20 & 11 & 9 & 9 && 137&1.0e3&6.6e3 \\
v150\_d30 & 8 & 6 & 6 && 1534&1.4e4&2.6e3 \\
v150\_d50 & 4 & 4 & 4 && 2&19&3.1e3 \\
v150\_d70 & 3 & 3 & 3 && 7&86&4.8e3 \\
v200\_d5 & 40 & \underline{30} & 29 && 5609&4.8e4&2.7e4 \\
v200\_d10 & 20 & \underline{17} & 16 && 10981&1.0e5&1.7e5 \\
v200\_d20 & 12 & 9 & 9 && 4513&5.6e4&3.9e5 \\
v200\_d30 & 8 & 7 & 7 && 19&2.7e2&1.6e5 \\
v200\_d50 & 5 & 4 & 4 && 141&2.6e3&8.4e3 \\
v200\_d70 & 3 & 3 & 3 && 1&23&9.6e3\\ 
\bottomrule
\end{tabularx}
\end{table*}

\begin{table*}[htb]
\scriptsize
\center
\caption{\label{table:ComGRASPMIP22CDS}Comparison of the  $MIP$ method from \cite{buchanan2015integer}, to the proposed greedy algorithm combined with the correction procedure $GrC$ and  the GRASP for the 2-2-CDS. "-" is used for instances without a feasible solution and "*" in case $GrC$ did not generate it.  Underlined values indicate that GRASP did not find optimal solution.}
\begin{tabularx}{350pt}{X*{8}{c}}
\toprule
Dataset& \multicolumn{3}{c}{$Solution$}&\hspace{50pt} & GRASP Iter. &\multicolumn{2}{c}{$Time[ms]$} & \\
		\cmidrule(r){2-4}\cmidrule(r){7-8}
       &   GrC& GRASP& MIP&& &GRASP &MIP \\
\midrule
v30\_d30  &9 & 8 & 8 && 2 & 2 & 60\\
v30\_d50  &5 & 5 & 5 && 2 & 2 & 0\\
v30\_d70  &4 & 4 & 4 && 1 & 1 & 30\\
v50\_d10  &* & 22 & 22& & 2 & 4 & 0\\
v50\_d20  &14 & 12 & 12 && 4 & 12 & 80\\
v50\_d30  &9 & 8 & 8 && 47 & 1.6e2 & 60\\
v50\_d50  &7 & 5 & 5 && 11 & 39 & 50\\
v50\_d70  &4 & 4 & 4 && 2 & 5 & 90\\
v70\_d5  &48 & 47 & 47 && 699 & 2.5e3 & 0\\
v70\_d10 &28 & 24 & 24 &&993 & 4.8e3 & 90\\
v70\_d20 &15 & 12 & 12 && 9229 & 4.8e4 & 1.4e2\\
v70\_d30  &9 & 8 & 8 && 220 & 1.2e3 & 2.0e2\\
v70\_d50  &* & 5 & 5 && 58 & 2.7e2 & 2.0e2\\
v70\_d70  &5 & 4 & 4 & & 1 & 3 & 3.6e2\\
v100\_d5  &46 & 44 & 44 && 19862 & 1.5e5 & 80\\
v100\_d10  &29 & \underline{25} & 24 && 729 & 5.9e3 & 4.8e2\\
v100\_d20  &16 & 13 & 13 && 511 & 4.1e3 & 4.5e2\\
v100\_d30  &11 & 9 & 9 && 18857 & 1.6e5 & 1.9e3\\
v100\_d50  &7 & 6 & 6 && 88 & 8.3e2 & 6.2e2\\
v100\_d70  &4 & 4 & 4 && 18 & 1.3e2 & 5.8e2\\
v120\_d10  &31 & \underline{25} & 24 && 563 & 5.6e3 & 9.8e2\\
v120\_d20  &17 & \underline{14} & 13 && 506 & 5.3e3 & 2.5e3\\
v120\_d30  &12 & 9 & 9 && 21824 & 2.4e5 & 3.0e3\\
v120\_d50  &7 & 6 & 6 && 31 & 3.9e2 & 1.8e3\\
v120\_d70  &5 & 4 & 4 && 9 & 90 & 1.9e3\\
v150\_d5  &* & \underline{48} & 45 && 21037 & 2.7e5 & 3.4e2\\
v150\_d10 &31 & \underline{27} & 24 && 2042 & 2.8e4 & 6.2e3\\
v150\_d20  &17 & \underline{15} & 13 && 135 & 1.9e3 & 2.8e4\\
v150\_d30  &10 & \underline{10} & 9 && 102 & 1.7e3 & 1.4e4\\
v150\_d50  &8 & 6 & 6 && 3.8e3 & 198 & 3.2e3\\
v150\_d70  &5 & 4 & 4 && 7.3e3 & 391 & 4.9e3\\
v200\_d5  &56 & \underline{54} & 48 && 6785 & 1.3e5 & 9.4e3\\
v200\_d10  &33 & \underline{29} & 26 && 7335 & 1.5e5 & 2.1e6\\
v200\_d20  &19 & \underline{16} & 14 & & 29 & 7.3e2 & 8.3e5\\
v200\_d30  &13 & \underline{11} & 10 && 98 & 2.7e3 & 3.2e5\\
v200\_d50  &8 & 6 & 6 && 13483 & 4.4e5 & 3.3e4\\
v200\_d70  &5 & \underline{5} & 4 && 8 & 2.9e2 & 3.6e4\\
\bottomrule
\end{tabularx}
\end{table*}

The results of the comparison are presented in Tables \ref{table:ComGRASPMIP21CDS},\ref{table:ComGRASPMIP22CDS}. In this table, we show the number of nodes in the best found dominating set for the non-randomized  greedy algorithm combined with the correction procedure ($GrC$) and the GRASP extension. In the case of GRASP a maximum of 25000 solutions was generated and we show the computational time, and corresponding iteration, for finding the best solution. In the case of $GrC$ execution times are not included as they are approximately equal to computational time divided by the iteration count for GRASP. The two methods are compared with the MIP method from \cite{Buchanan2014410}. For both the GRASP and $GrC$ the maximal size of the candidate list is $M=500$. In case of GRASP the value of the randomization parameter  is $\alpha_0 = 1.25$.  The node $u$ having the highest number of adjacent nodes was used as the root node of the $GrC$. 

In the case of the 2-1-CDS, GRASP manages to find optimal solutions in 37 out 40 problem instances. The instances for which it does not manage to find optimal solutions are large sparse graphs. For this type of graphs, the computational time is worse than $MIP$. In general, GRASP had significantly lower computational times than MIP for highly dense graphs. $GrC$ had very short execution times and even for the largest problem instances it is less than 25 milliseconds, and manages to find optimal solutions for 15 problem instances.  The GRASP algorithm seems slightly less effective in the case of 2-2-CDS, when the quality of solutions is considered, in which it finds  27 optimal solutions out of 38 tested graphs. The computational advantage to MIP for highly dense graphs is much lower compared to  2-1-CDS due to a large number of generated solutions. In case of 2-2-CDS, $GrC$ is also very fast taking less than 40 milliseconds to find solutions for the largest problem instances. Although $GrC$ manages to generates good approximate solutions it only manages to find 4 optimal solutions. This illustrates the effectiveness of using the randomization in GRASP. It is important to note that $GrC$ does not find feasible solutions in $3$ cases due to the exclusion of candidate ears not improving the partial solution as mentioned in Section \ref{GreSec}.

 \section{Conclusion}
In this paper, we have presented a GRASP algorithm for solving the 2-$m$-CDS. The algorithm is based on greedily growing a 2-connected subgraph by extending it with open ears selected using a heuristic function. Further, the quality of solutions generated in this way is improved using a simple correction procedure and adding randomization. Our computational results show that the proposed method is highly competitive with recently published MIP based algorithm. It is important to note that in general heuristic and GRASP based algorithms are designed for finding high quality approximate solutions and not the best choice for finding optimal ones but even in such a comparison the proposed methods prove to be better than MIP for some specific types of graphs. We wish to emphasize that this is a novel approach for solving the 2-m-CDS that has potential for a wide range of potential applications and extensions.

For example, due to the fact that the proposed greedy algorithm manages to find good quality approximate solutions in a very short computational time, it can be used for providing upper bounds for MIP models. Another potential avenue of research is adapting the proposed algorithm to more complex metaheuristics like for instance the ant colony optimization. In the future, we plan to extend the proposed method to a matheuristic paradigm by developing a MIP based local search. Further, we plan to explore adapting the presented method to similar problems like the 2-connected vertex cover problem.


%
%


\end{document}